\documentclass[twocolumn,           
showpacs,            
               preprintnumbers,     
               aps,                 
               prd,          	
               a4paper,             
               superscriptaddress,  
               nofootinbib,         
               tightenlines,        
               floats               
               ]{revtex4}

\usepackage{graphicx}
\usepackage{amssymb,latexsym}

 \newcommand{\CF}{{\cal F}}

 \newcommand{\CL}{{\cal L}}
 \newcommand{\CN}{{\cal N}}
 \newcommand{\calP}{{\cal P}}
 \newcommand{\CR}{{\cal R}}

\newcommand{\bear}{\begin{array}}  \newcommand{\eear}{\end{array}}
\newcommand{\bea}{\begin{eqnarray}}  \newcommand{\eea}{\end{eqnarray}}
\newcommand{\beq}{\begin{equation}}  \newcommand{\eeq}{\end{equation}}
\newcommand{\bef}{\begin{figure}}  \newcommand{\eef}{\end{figure}}
\newcommand{\bec}{\begin{center}}  \newcommand{\eec}{\end{center}}
  
\newcommand{\lmk}{\left(}  \newcommand{\rmk}{\right)}
\newcommand{\lkk}{\left[}  \newcommand{\rkk}{\right]}

\newcommand{\la}{\left\langle} \newcommand{\ra}{\right\rangle}

\begin{document}

\title{ Band-power reconstruction of the primordial fluctuation spectrum \\
by the maximum likelihood reconstruction method}


\author{Ryo Nagata}
\affiliation{Research Center for the Early Universe (RESCEU),
Graduate School of Science, The University of Tokyo, Tokyo 113-0033,
Japan }
\author{Jun'ichi Yokoyama}
\affiliation{Research Center for the Early Universe (RESCEU),
Graduate School of Science, The University of Tokyo, Tokyo 113-0033,
Japan }
\affiliation{Institute for the Physics and Mathematics of the Universe(IPMU),
The University of Tokyo, Kashiwa, Chiba, 277-8568, Japan}

\date{\today} 


\preprint{RESCEU-64/08}


\begin{abstract}
The primordial curvature fluctuation spectrum is reconstructed 
by the maximum likelihood reconstruction method using the five-year 
Wilkinson Microwave Anisotropy Probe data 
of the cosmic microwave background temperature anisotropy. 
We apply the covariance matrix analysis and decompose the reconstructed spectrum 
into statistically independent band powers. 
The prominent peak off a simple power-law spectrum found in our previous analysis turn out to be a $3.3\sigma$ deviation. 
From the statistics of primordial spectra reconstructed from mock observations, 
the probability that a primordial spectrum including such excess is realized in a power-law model 
is estimated to be about $2$\%. 
\end{abstract}

\maketitle

\section{Introduction}

The cosmic evolution during the inflationary stage of 
the early Universe \cite{inf1,inf2,inf2a,inf3} 
is recorded in the primordial fluctuation spectrum \cite{yuragi,yuragi2,yuragi3},  
which can be revealed by high-precision observations of 
the cosmic microwave background (CMB) and large-scale structures. 
If we found some nontrivial features exceeding a level 
expected by the cosmic variance after proper analysis of 
observational errors, we would have to seriously consider its generation 
mechanism, namely, an inflation model that generates such 
a nontrivial fluctuation spectrum. This would not only be 
a challenge to particle-physics model building but also a boon 
that provides an important clue to high-energy physics. 
The CMB anisotropy data provided by the Wilkinson Microwave Anisotropy Probe (WMAP) mission 
\cite{WMAPBASIC,WMAPTEMP,WMAPCOSMO,WMAPINF,WMAPTTTE,loglike,WMAP3TEMP,WMAP5,WMAP5ONLY,WMAP5COSMO} 
is already on the precision level sufficient for such purpose. 

There have been persistent controversy on the existence of nontrivial features in 
the CMB temperature anisotropy spectrum. 
The angular spectrum of the first-year WMAP data \cite{WMAPBASIC,WMAPTEMP,WMAPCOSMO,WMAPINF,WMAPTTTE} 
exhibited a sign of a running spectral index, oscillatory behaviors on intermediate scales, 
and lack of power on large scales, 
which cannot be explained by a power-law primordial spectrum 
that is a generic prediction of simplest inflation models. 
These features may provide clues to unnoticed physics of inflation. 
Some of these anomalous structures disappeared on the three-year spectrum, 
but several anomalies are still remaining \cite{WMAP3TEMP}. 
To explain these features, a number of inflation models have been proposed 
\cite{CPKL03,CCL03,FZ03,KT03,HM04,LMMR04,KYY03,YY03,KS03,YY04,MR04A,MR04C,KTT04,HS04,HS07,CHMSS06}. 

To reconstruct the primordial spectrum 
using CMB anisotropy data without specific prior assumptions about the shape of the primordial spectrum, 
various observational methods have been proposed 
\cite{WSS99,BLWE03,MW03A,MW03B,MW03C,MW03D,SH01,SH04,BLH07,SVJ05,WMAP3COSMO,LSV08,VP08}. 
Among such attempts, nonparametric methods \cite{MSY02,MSY03,KMSY04,KSY04,KSY05,NY08,SS03,TDS04,THS05,SS06,SS07}, 
which can reconstruct the primordial spectrum as a continuous function without any {\it ad hoc} filtering scale, 
have a merit for investigating a detailed feature localized in a narrow wave number range. 
Recently, we reconstructed the primordial spectrum by the cosmic inversion method 
\cite{MSY02,MSY03,KMSY04,KSY04,KSY05,NY08} 
from the temperature anisotropy spectrum of the five-year WMAP data. 
The reconstructed spectrum accurately preserved the information of detailed features observed on the angular spectrum. 
To distinguish possible true signal from the cosmic variance appropriately, 
we applied the covariance matrix analysis and decomposed the reconstructed spectrum 
into statistically independent band powers. 
Although the reconstructed spectrum was basically consistent with a simple power-law spectrum, 
it exhibited anomalous deviation localized on $\approx 700$Mpc scale, which was unfortunately 
near the border of the wave number domain where accurate reconstruction was possible by the cosmic inversion method. 
To investigate the possible fine structure, we need to adopt another reconstruction method 
which can probe a detailed feature of the primordial spectrum in the corresponding wave number range. 

The purpose of this work is to revisit the possibility of fine structure in the primordial spectrum. 
We apply the nonparametric reconstruction method, which is originally developed by  
Tocchini-Valentini et al.\cite{TDS04,THS05}, to 
the five-year WMAP temperature anisotropy spectrum \cite{WMAP5} and perform the band-power 
analysis introduced in our previous work. 
Because of the arbitrariness of the primordial spectrum, we inevitably incorporate 
infinite degree of freedom to our analysis, which results in degeneracy 
among spectral shape and cosmological parameters \cite{KNN01,SBKET,SS07}. 
However, the fine structure on which we focus in this paper is much narrower 
than the structure dependent on the conventional cosmological parameters whose 
characteristic width is principally the inverse of the horizon length at the recombination epoch. 
As shown in \cite{KMSY04}, different choices of cosmological parameters affect only the overall shape and 
the fine structures of the reconstructed spectrum remain intact. 
In this paper, we consider the concordance adiabatic $\Lambda$CDM model, 
where the cosmological parameters (except for the ones characterizing the primordial spectrum) 
are those of the WMAP team's best-fit power-law model \cite{WMAP5ONLY,WMAP5COSMO}, 
and instead focus on the detailed shape of the primordial spectrum. 

This paper is organized as follows: 
In Sec. II, the overview of our analysis is described. 
In Sec. III, we show the reconstructed primordial power spectrum 
from the five-year WMAP data and discuss its implication. 
Finally, Sec. IV is devoted to the conclusion.

\section{Method}

\subsection{Reconstruction formula}
We introduce the maximum likelihood reconstruction method\footnote{This method was 
called as {\it nonparametric reconstruction} by the authors of the preceding works. 
We compare the result of reconstruction with that by another 
non-parametric method, namely the cosmic inversion method. Therefore, 
we call it the maximum likelihood reconstruction method for distinction in this paper.} 
to reproduce the primordial power spectrum $P(k)$. 
Temperature anisotropy is decomposed into the coefficients of spherical harmonics as, 
\begin{eqnarray}
\frac{\delta T}{T}(\hat{n}) = \sum_{\ell,m} a_{\ell m} Y_{\ell m}(\hat{n}) .
\end{eqnarray}
A theoretical angular power spectrum $C_\ell$ is the ensemble average of their norm 
which is related to $P(k)$ via a radiation transfer function $X_\ell (k)$, 
\begin{eqnarray}
C_\ell = \langle |a_{\ell m}|^2 \rangle 
 = \frac{2}{\pi} \int dlnk \hspace{0.1cm} k^3P(k) \lmk \frac{X_\ell (k)}{2\ell+1} \rmk^2 .
\label{CLen}
\end{eqnarray}
We define the primordial spectrum as the initial spectrum of curvature fluctuation, 
$P(k)= \langle |\CR(0,k)|^2 \rangle$. 
The probability distribution of a harmonic coefficient for a given $P(k)$ obeys 
to Gaussian statistics of a complex variable, 
\begin{eqnarray}
\calP[a_{\ell m}|P(k)] &=& \frac{1}{\pi C_\ell}\exp\lmk - 
\frac{|a_{\ell m}|^2}{C_\ell}\rmk \hspace{0.5cm} (m \neq 0), \\
\calP[a_{\ell 0}\hspace{0.1cm}|P(k)] &=& \frac{1}{\sqrt{2\pi C_\ell}}\exp\lmk - 
\frac{|a_{\ell 0}|^2}{2C_\ell}\rmk. 
\label{CLpr}
\end{eqnarray}
Hence the probability of realizing a sky (i.e. a set of harmonic coefficients) is the product of them. 
\begin{eqnarray}
 \calP[\{a_{\ell m}\}|P(k)] = \prod_{\ell, \hspace{0.1cm}m\ge0} \calP[a_{\ell m}|P(k)].
\label{CLpr}
\end{eqnarray}
Defining 
\beq
C_\ell^{obs} \equiv \frac{1}{2\ell +1}\sum_{m=-\ell}^\ell
 |a_{\ell m}|^2,
\eeq
we find the following log-likelihood function as 
\begin{eqnarray}
 \CL &=& -\ln \calP[\{C_\ell^{obs}\}|P(k)] \\
     &=& \sum_\ell
\frac{(2\ell+1)}{2} \lkk \frac{\hspace{0.3cm} C_\ell^{obs}}{C_\ell}
+\ln C_\ell \rkk + (const.) \hspace{0.1cm}.
\end{eqnarray}
If Gaussian distributed measurement noise associates with observed $a_{\ell m}$, 
the expected mean of the observed angular power spectrum increases due to the noise contribution. 
Then, the above equations should be modified as 
\begin{eqnarray}
C_\ell^{obs} \equiv \frac{1}{2\ell +1}\sum_{m=-\ell}^\ell
 |a_{\ell m}|^2 - N_\ell, 
\end{eqnarray}
and
\begin{eqnarray}
 \CL = \sum_\ell
\frac{ (2\ell+1) }{2} \lkk \frac{ C_\ell^{obs}+N_\ell}{ \hspace{0.3cm} C_\ell+N_\ell}
+\ln(C_\ell+N_\ell) \rkk \nonumber \\ + (const.) \hspace{0.1cm}.
\end{eqnarray}
Here we parametrize such noise contribution by $N_\ell$. 
When all-sky information is not available, 
the above log-likelihood function scales with a sky coverage fraction $f_{sky}$ \cite{LV}. 

In this method, 
we would like to obtain a power spectrum $P(k)$, which maximizes 
the above likelihood functional for a given observation. 
Our master equation consists of
 a functional derivative of $\CL$ with respect to $P(k)$.
\beq
\frac{\delta\CL}{\delta P(k)}=\sum_\ell
f_{sky}
\frac{k^2}{\pi}\frac{|X_\ell(k)|^2}{2\ell+1} 
\frac{C_\ell-C_\ell^{obs}}{(C_\ell+N_\ell)^2} 
=0,  \label{master}
\eeq
where we have used
\beq
 \frac{\delta C_\ell}{\delta P(k)}
=\frac{2 k^2}{\pi}\frac{|X_\ell(k)|^2}{(2\ell+1)^2}.
\eeq
One can rewrite (\ref{master}) as
\begin{eqnarray}
  \sum_\ell \frac{k^2 \hspace{0.1cm} f_{sky}}{\pi (C_\ell+N_\ell)^2}
\frac{|X_\ell(k)|^2}{2\ell+1}
\int dk' \frac{2k'^2}{\pi}\frac{|X_\ell(k')|^2}{(2\ell+1)^2}P(k') \nonumber \\
= \sum_\ell \frac{k^2 \hspace{0.1cm} f_{sky}}{\pi (C_\ell+N_\ell)^2}
\frac{|X_\ell(k)|^2}{2\ell+1} C_\ell^{obs}.  \hspace{1.0cm} \label{master2}
\end{eqnarray}
Discretizing in $k'$ and introducing matrices
\bea
  D_{k\ell}&\equiv&\frac{k^2 \hspace{0.1cm} f_{sky}}{\pi (C_\ell+N_\ell)^2}
\frac{|X_\ell(k)|^2}{2\ell+1}, \label{matD} \\
  G_{\ell k'}&\equiv& 
\frac{2k'^2}{\pi}\frac{|X_\ell(k')|^2}{(2\ell+1)^2} \Delta k',
\eea
and vectors $P_{k'}\equiv P(k')$ and $C_\ell^{obs}$,
(\ref{master2}) can be recast into a matrix equation,
\beq
  \sum_{\ell, \hspace{0.1cm} k'} D_{k\ell}G_{\ell k'}P_{k'}= \sum_\ell D_{k\ell}C_\ell^{obs}. 
\label{matrix}
\eeq
Although the matrix $DG$ is a square matrix with its dimension
equal to the number of discretized points in $k$, we cannot
invert it because the mapping from $P(k)$ to $C_\ell$ is not one to one. 
Operation of the matrix $G$ erases two kinds of information in $P$. 
One is fine structure in $k$ space, which is smoothed by the convolution with the transfer function. 
The other is power on the scales of acoustic troughs where the transfer function is almost vanishing. 
Hence the actual reconstruction suffers a resolution limit and computational noise on specific scales. 

According to the Bayes theorem,
the conditional probability of $P(k)$ under the condition that each $a_{\ell m}$
takes some observed value reads,
\beq
  \calP \lkk P(k)|\{C_\ell^{obs}\}\rkk=
\frac{\calP\lkk\{C_\ell^{obs}\}|P(k)\rkk
\calP\lkk P(k)\rkk}
{\calP\lkk\{C_\ell^{obs}\}\rkk}.
\eeq
Here, following the preceding works \cite{TDS04,THS05}, 
we assume that $P(k)$ should be a sufficiently smooth function. 
As the prior probability for $P(k)$, we adopt
\beq
 \calP\lkk P(k)\rkk \propto \exp\lkk -\epsilon \int dk \hspace{0.1cm} 
 \frac{1}{2} \lmk\frac{dk^3P(k)}{d k}\rmk^2\rkk 
 \equiv e^{ -\epsilon \CR [P(k)]},
\eeq
where $\epsilon$ is a prior parameter. 
Then, Eq.(\ref{master}) is modified to
\beq
  \frac{\delta~}{\delta P(k)}\Big( \CL[P(k)]+\epsilon \CR[P(k)]\Big)=0,
\eeq
and the matrix Eq. (\ref{matrix}) is modified accordingly to
acquire $\epsilon$ dependence. 
As mentioned above, there can be many possible $P(k)$ whose resultant $C_\ell$ are indistinguishable. 
Now we have introduced the guiding principle which favors smoother $P(k)$. 
We can interpret $\CL[P(k)]+\epsilon \CR[P(k)]$ as the action of a forced oscillator 
rolling around $C_\ell^{obs}$, which assures that the reconstructed $P(k)$ restores the observation. 
We perform the reconstruction procedure 
by solving the discretized equation for $P(k)$ as 
\begin{eqnarray}
\sum_{\ell, \hspace{0.1cm} k', \hspace{0.1cm}k''} 
\lmk D_{k\ell}G_{\ell k'} + \epsilon \hspace{0.05cm} k^3 \hspace{0.15cm}
{}^t\CF_{k \hspace{0.1em} k''} \CF_{k'' k'} \hspace{0.05cm} {k'}^3 \rmk P_{k'} \nonumber \\
= \sum_\ell D_{k\ell}C_\ell^{obs}, 
\label{invert}
\end{eqnarray}
where the first derivative matrix $\CF$ gives the difference between the neighboring elements of an operand 
over the discretization width and $^t\CF$ denotes its transposed matrix which acts like $-\frac{d}{dk}$. 
Practically such discretization results in constraining the first derivatives 
at both boundaries of the $k$ range concerned, which does not affect the results of reconstruction 
as long as the reconstruction is performed with sufficiently large buffers around the boundaries.

\subsection{Numerical calculation}
We adopt the adiabatic initial condition and fiducial cosmological 
parameters found by the WMAP team \cite{WMAP5COSMO} to calculate the transfer function. 
Specifically, the cosmological parameters are $h=0.724$, 
$\Omega_b=0.0432$, $\Omega_\Lambda=0.751$, $\Omega_m=0.249$, and $\tau=0.089$, 
where the distance to the last scattering surface is $d \simeq 1.43\times 10^4$Mpc. 
The transfer function is calculated by the routines of CMBFAST code \cite{SZ96} 
with the improved resolution sufficiently for reconstruction of fine structure. 
Hereafter, we treat $A(k) \equiv k^3P(k)$ instead of $P(k)$ itself for the comprehensive display purpose 
and consistency with the common normalization of fluctuation amplitude.

We focus on the angular scales of $30 \lesssim \ell \lesssim 390$ because 
the actual likelihood evaluation on the largest angular scales is rather complicated and 
the observational data is noise dominated on the scales of $\ell > 400$. 
Since $C_\ell$ mainly depends on the $k$ modes spreading around $k=\ell/d$, 
the corresponding wave number range is 
$ 2.10 \times 10^{-3}$ Mpc${}^{-1} \lesssim k \lesssim 2.73 \times 10^{-2}$ Mpc${}^{-1}$. 
To avoid the boundary effect, we incorporate the angular spectrum data 
of $10 \leq \ell \leq 800$ and perform the reconstruction in the 
much wider range of $1.13 \times 10^{-5}$ Mpc${}^{-1} \lesssim k \lesssim 2.06 \times 10^{-1}$ Mpc${}^{-1}$, 
which is divided into about $6000$ bins. 

A part of the matrix $D$ defined in Eq.(\ref{matD}) corresponds to the Fisher matrix for the angular spectrum data. 
Now we identify the factor, $\frac{(2\ell+1)f_{sky}}{2(C_\ell+N_\ell)^2}$, 
with the diagonal elements of the Fisher matrix provided by the WMAP team. 
In the actual Fisher matrix, the diagonal elements have the additional $f_{sky}$ factor 
and the off-diagonal elements are not vanishing 
because of the correlation which comes from incompleteness of the observational sky coverage \cite{loglike}. 
It is straightforward to incorporate these aspects to our formalism. 
Substituting the Fisher matrix itself, $N^{-1}$, to the factor, $\frac{(2\ell+1)f_{sky}}{2(C_\ell+N_\ell)^2}$, 
we redefine the matrix $D$ as follows:
\bea
  D_{k\ell} &\equiv& \sum_{\ell'} \lmk N^{-1} \rmk_{\ell\ell'}\frac{2k^2}{\pi} \frac{|X_{\ell'}(k)|^2}{(2\ell'+1)^2}.
\eea
We take account of the entire Fisher matrix in the actual reconstruction. 
Note also that the matrix $D$ depends on $C_\ell$, which is a functional of $A(k)$. 
When we invert the matrix Eq. (\ref{invert}), we first put 
$C_\ell^{pl}$, which is the angular power 
spectrum calculated from the best-fit power-law primordial spectrum. Once a solution 
for $A(k)$ is thereby obtained, we calculate $C_\ell$ from it and 
insert the new $C_\ell$ to the matrix $D$ to repeat the matrix
inversion. We thus solve for $A(k)$ iteratively. 
The solution relaxes to the same form after first iteration because fluctuation around the mean curve of $C_\ell$ 
does not seriously affect the matrix $D$. 

Before proceeding to the statistical treatment, let us discuss the criterion for choosing the value of $\epsilon$. 
Equation (\ref{invert}) indicates that the result of reconstruction depends on the parameter $\epsilon$. 
As shown in the next section, the reconstructed primordial spectrum has oscillatory modulation. 
The locations of those peaks are almost independent of $\epsilon$, 
while the amplitude of the peaks depends on $\epsilon$. 
For instance, in the preceding works, the value of $\epsilon$ is chosen so that 
the reduced $\chi^2$ value of the restored angular spectrum from the reconstructed primordial 
spectrum equals unity \cite{TDS04,THS05}. 
Here, it should be noted that a maximum likelihood solution does not necessarily coincide with an 
underlying primordial spectrum. 
Indeed, a reconstructed spectrum from a simulated observational angular spectrum based on some 
power-law primordial spectrum usually has oscillatory modulation, which comes from the cosmic variance, 
even if we chose the value of $\epsilon$ 
so that the reduced $\chi^2$ value of the restored angular spectrum is unity. 
In regard to signal detection, however, the most important quantity is not the amplitude itself but the ratio of 
the signal amplitude to the error magnitude. 
Robustness of a statistical conclusion against changing the value of $\epsilon$ is crucial to our reconstruction scheme. 
Through the error analysis described in the next subsection, we find that 
the statistical error of the peak amplitude depends on $\epsilon$ in the same manner as the peak amplitude itself 
and the different choice of 
$\epsilon$ does not seriously affect the statistical significance of the reconstructed features. 
We set the value of $\epsilon$ to $4 \times 10^{-4}$Mpc$^{-1}$ so that 
the power of the reconstructed primordial spectrum is positive on any scale and 
the reconstructed primordial spectrum fits the result of our previous work 
in some restricted wave number range, which 
enables to directly compare the results of two different reconstruction methods. 
The further discussion is described in the appendix. 

\subsection{Monte Carlo simulation}\label{wind}
It is theoretically expected that different wave number modes 
of primordial fluctuation are statistically independent with each other. 
On the other hand, neighboring wave number modes 
in the observationally reconstructed primordial spectrum are highly 
correlated with each other because of the finiteness of the 
width of the transfer function. Hence in order to perform 
proper statistical analysis of the reconstructed spectrum, 
it is important to decompose the reconstructed power spectrum into 
mutually independent band powers. 

For the above purpose, we employ the Monte Carlo method to calculate the covariance matrix of the 
reconstructed power spectrum. 
Producing $50000$ realizations of a temperature anisotropy spectrum based on the WMAP team's best-fit power-law model 
whose statistics obey the likelihood function provided by WMAP team with good precision, 
we obtain $50000$ realizations of a reconstructed primordial spectrum. 
(The detailed prescription of generating mock anisotropy spectra is explained in our previous paper \cite{NY08}.)
The covariance matrix of the reconstructed power spectrum is defined by 
\bea
 K_{ij}&\equiv& \frac{1}{\CN}\sum_{\alpha=1}^{\CN} A_\alpha(k_i)
 A_\alpha(k_j)- \frac{1}{\CN}\sum_{\alpha=1}^{\CN} A_\alpha(k_i)
\frac{1}{\CN}\sum_{\beta=1}^{\CN} A_\beta(k_j) \nonumber \\
&\equiv& \la\!\la A_\alpha(k_i)A_\alpha(k_j)\ra\!\ra_{\alpha}
-\la\!\la A_\alpha(k_i)\ra\!\ra_{\alpha}\la\!\la A_\beta(k_j)\ra\!\ra_{\beta},
\eea
where $A_\alpha(k_i)$ represents the value of the reconstructed power
spectrum at $k=k_i$ in the $\alpha$-th realization, and $\CN=50000$
as mentioned above. 

Significant correlation can be seen between the neighboring bins in the covariance matrix 
and the characteristic correlation width is $\sim 10/d$ (see Sec.\ref{BPA}). 
To extract the mutually independent degree of freedom and estimate the reliable error bars, 
we disentangle this correlation by diagonalizing the covariance matrix. 
Since $K$ is a real symmetric matrix, it can be diagonalized by a
real unitary matrix $U$, to yield
\beq
  UKU^\dag = {\rm diag}\lmk \lambda_1, \lambda_2,...,
  \lambda_N\rmk\equiv \Lambda,
\eeq
where $\lambda_i$'s are the eigenvalues of $K$. We find that they are
positive definite as they should be, provided that
 we take $N$ small enough so that neighboring
modes are not degenerate with each other.  In the present case,
we find that if we take $N\lesssim 100$, the covariance matrix
is well behaved in the sense that the following procedure is
possible with positive definite $\lambda_i$ and the well-behaved
window matrix  defined below. 
In terms of 
\beq
\Lambda^{-1/2}\equiv
{\rm diag}\lmk \lambda_1^{-1/2}, \lambda_2^{-1/2},...,
  \lambda_N^{-1/2}\rmk ,
\eeq
 we define inverse square root of $K$
as $K^{-1/2}\equiv U^\dag \Lambda^{-1/2}U$.  
We also define a
window matrix $W$ by
\beq
  W_{ij}=\frac{(K^{-1/2})_{ij}}{\sum_{m=1}^{N}(K^{-1/2})_{im}},
\eeq
which satisfies the normalization condition $\sum_{j=1}^{N}W_{ij}=1$.
Convolving $A_\alpha(k_i)$ with this window function, we define a new
statistical variable $S_\alpha(k_i)$ as
\beq
  S_\alpha(k_i)\equiv \sum_{j=1}^N W_{ij}A_\alpha(k_j),
\eeq
whose correlation matrix is diagonal and reads
\bea
  \la\!\la S_\alpha(k_i)S_\alpha(k_j)\ra\!\ra_{\alpha}
-\la\!\la S_\alpha(k_i)\ra\!\ra_{\alpha}\la\!\la 
S_\beta(k_j)\ra\!\ra_{\beta} \nonumber \\
= (WK^tW)_{ij}
 =\lkk \sum_{m=1}^N (K^{-1/2})_{im}\rkk^{-2}\delta_{ij}, 
\eea
where $^tW$ denotes a transposed matrix.

\begin{figure}
\includegraphics[scale=1.1]{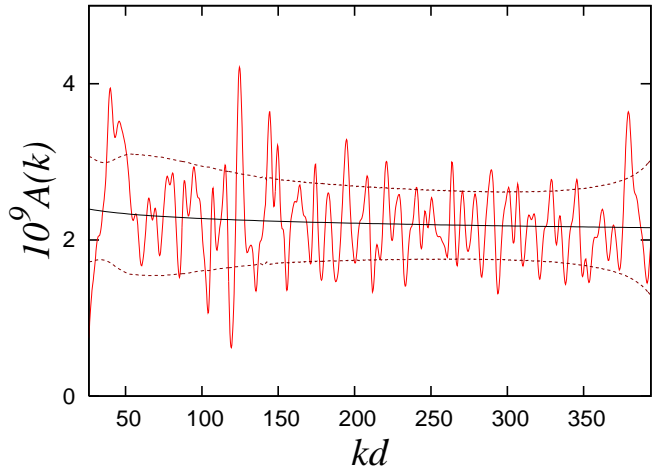}
\caption{
The reconstruction of the primordial spectrum by the maximum likelihood reconstruction method from 
the five-year WMAP data.  The solid wavy curve is the result of reconstruction
 and the straight line is the best-fit power-law spectrum.  Dotted
 lines are the standard deviation around the best-fit power law.
}
\label{fig:rec}
\end{figure}

\begin{figure}
\includegraphics[scale=1.1]{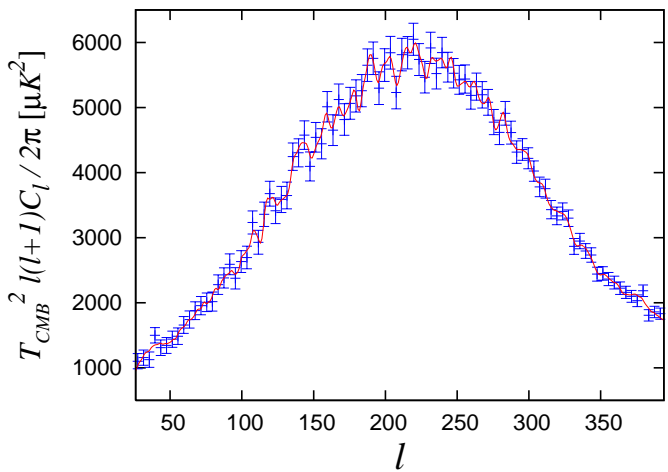}
\caption{
The comparison of the binned five-year WMAP data of $\Delta \ell = 4$ 
with the anisotropy spectrum restored from the reconstructed primordial spectrum. 
}
\label{fig:clbin}
\end{figure}

\section{Results of reconstruction}

\subsection{Reconstructed primordial spectrum}
Figure\ \ref{fig:rec} shows the reconstructed primordial spectrum from the five-year WMAP data and 
associated $1\sigma$ standard errors which correspond to the diagonal elements of 
the error covariance matrix (see Sec. \ref{BPA}) calculated by Monte Carlo simulation. 
In this figure, the solid straight line is the best-fit power-law 
$\Lambda$CDM model obtained by the five-year WMAP observations, namely, 
the power-law spectrum with $A=2.39\times 10^{-9}$ and $n_s=0.961$, 
where $A$ is the amplitude of curvature perturbation at $k_0=0.002{\rm Mpc}^{-1}$. 

The reconstructed spectrum has the rapid oscillations whose amplitude is about $20\%-30\%$ of the mean value 
because the reconstruction maps also the cosmic variance on the anisotropy spectrum into wave number space. 
The modulations of the reconstructed spectrum roughly fit inside the $1\sigma$ borders; 
therefore it is not necessarily required that the inflation model 
responsible for creation of our Universe should predict such a highly nontrivial spectrum. 

The reconstructed spectrum seems consistent with the power-law spectrum on most scales 
except for the large modulation around the scales of $kd \simeq 120$ (or equivalently $\approx 700$Mpc). 
In our previous analysis by the cosmic inversion method, 
this feature was estimated as about $4\sigma$ deviation from a power-law spectrum, 
however it was located near the border of the wave number domain where accurate reconstruction was possible. 
In this stage, it seems to be more moderate deviation. 
To evaluate the statistical significance of the reconstructed spectrum appropriately, 
we perform band-power decorrelation analysis. 
The result of our statistical analysis is described in Sec. \ref{BPA}. 

\subsection{Restored anisotropy spectrum}
Figure\ \ref{fig:clbin} illustrates the CMB temperature anisotropy spectrum, 
which we restore from the reconstructed primordial spectrum by adopting the cosmological parameters of 
the best-fit power-law model. 
The effective $\chi^2$ value in the range of $30 \le \ell \le 390$ for this restored anisotropy spectrum, 
which we calculate using the likelihood tool 
provided by the WMAP team \cite{LAMBDA}, is 327, 
while that for the best-fit power-law model is 408. 
Although the degree of fit is improved significantly,
it does not have the original statistical 
meaning because we have incorporated a functional degree of freedom. 

The restored anisotropy spectrum traces the fine structure of the observational spectrum fairly well.  
In particular, it fits the feature of the anomalous peak, which is located around $\ell\simeq120$. 
The maximum likelihood reconstruction method reproduces the phase of the observed fine structure 
more accurately than the cosmic inversion method and, as a result, 
the location of the anomaly in the reconstructed primordial spectrum has moved to 
slightly larger wave number ($kd = 125$) 
than the one evaluated by the cosmic inversion method (see the last paragraph of the appendix). 
The effective $\chi^2$ value in the range of $100 \le \ell \le 140$ is 42, 
while that for the best-fit power-law model is 61. 
In the angular spectrum, the anomaly does not look so prominent as in the primordial spectrum 
because it is located on the steep slope of the first Doppler peak. 

\begin{figure}
\includegraphics[height=8cm,width=8cm]{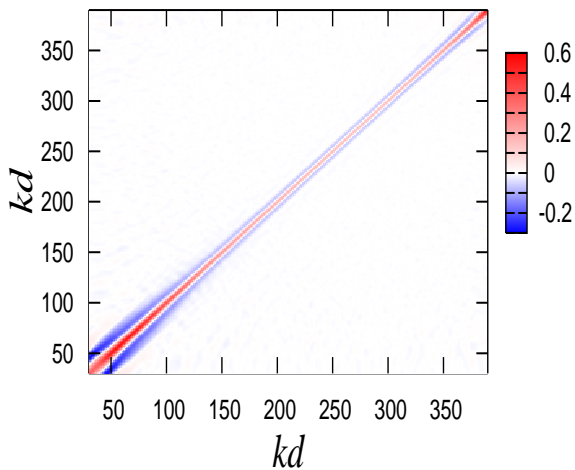}
\caption{
The error covariance matrix of the reconstructed primordial spectrum, $10^9A(k)$, calculated by Monte Carlo simulation. 
The central strip is indicating positive correlation 
and the surrounding strips are indicating negative correlation. 
Strong correlation on the largest scales comes from the curvature of the last scattering sphere 
while error regain on the small scale end of our analysis is due to the domination of measurement error. 
}
\label{fig:cov}
\end{figure}

\begin{figure}
\includegraphics[scale=1.1]{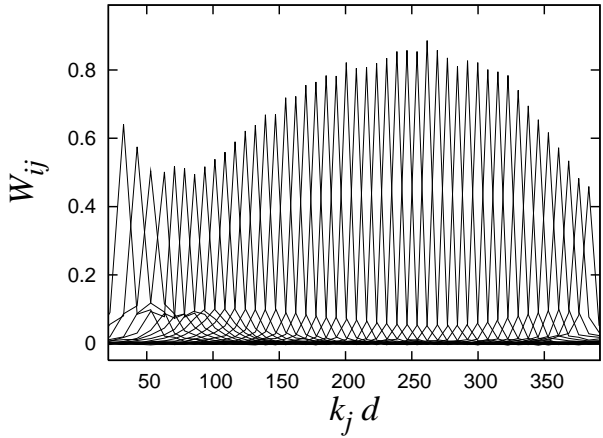}
\caption{The decorrelated window functions $W_{ij}$ for the intermediate $46$ band powers. 
About 20 modes corresponding to the inner most scales have higher peaks than those to surrounding scales 
whose strong correlation with neighboring $k$ modes broadens window functions. 
}
\label{fig:window}
\end{figure}

\begin{figure}
\includegraphics[scale=1.1]{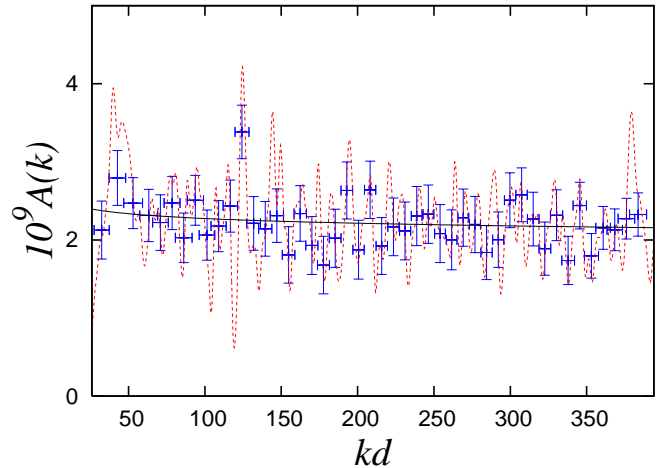}
\caption{
The band-power decomposition of the primordial spectrum reconstructed from the five-year WMAP data. 
Each data point indicates the amplitude of diagonalized mode $S$ defined in the text. 
The horizontal bar indicates the effective width of each window function 
which depicts the dispersion of the fitted Gaussian distribution.
}
\label{fig:band}
\end{figure}

\subsection{Band-power analysis} \label{BPA}
As we can see in the covariance matrix depicted in Fig.\ \ref{fig:cov}, 
neighboring wave number modes strongly correlate. The origin of correlation is 
the convolution with the transfer function. 
In particular, mapping the primordial spectrum to $C_{\ell}$ erases 
the information that is responsible for the fine structure whose characteristic scale is 
below the correlation width $\sim 10/d$. 
Although our reconstructed primordial spectrum restores the 
fine structure of the observed angular power spectrum well, 
the oscillations observed in the raw reconstructed spectrum may not have true statistical meaning. 

To extract statistically proper information, 
we decompose the power spectrum into mutually independent modes. 
Here, we construct band powers using the window matrix $W$ defined in
Sec. \ref{wind}, which diagonalizes the covariance matrix and gives
mutually independent errors. In order for appropriate sampling, 
we take the dimension of the window matrix so that $\Delta kd \simeq 8$. 
The bands corresponding to the wave number modes which lies between 
$kd = 30$ and $kd = 390$ are the intermediate $46$ bands. 
Figure \ref{fig:window} shows the window functions for each mode. 

Figure \ref{fig:band} is the result of band-power analysis of the five-year WMAP data. 
In this graph $i$-th data point indicates the value of 
\beq
 S(k_i) = \sum_{j=1}^N W_{ij} A(k_j),
\eeq
and the vertical error bar represents the variance
\begin{eqnarray}
\lkk\la\!\la S_\alpha^2(k_i)\ra\!\ra_{\alpha} -
\la\!\la S_\alpha(k_i)\ra\!\ra_{\alpha}^2\rkk^{1/2} \hspace{1.0cm} \nonumber \\
\hspace{1.0cm} = \lkk \sum_{m=1}^N (K^{-1/2})_{im}\rkk^{-1}.
\end{eqnarray}
Here, $k_i$ is the location of the peak of the $i$-th line of the window matrix $W_{ij}$. 
The horizontal bar, on the other hand, indicates the width of 
the window matrix, where the dispersion of the fitted Gaussian is shown. 
We find their typical width is $\Delta kd \approx 10$ which is 
basically determined by the width of the transfer function. 

The $i$-th band power depends on the multipole moment on the relevant angular scale of $\ell \sim k_id$ and 
also on the surrounding multipoles of $\ell \sim k_id \pm 5$. 
Indeed, we found that the statistics of each band-power subject to Gaussian distribution 
due to the superposition of several multipole moments 
even though the distribution of simulated $C_\ell$ is non-Gaussian. 
By virtue of our band-power analysis, we can also estimate the 
statistical significance of the reconstructed spectrum itself by evaluating the 
deviation from the best-fit power-law spectrum at every band simultaneously. 
We estimated the whole statistical significance of the 46 bands, which correspond to the scales of 
$30 \lesssim kd \lesssim 390$, and found that, 
in terms of reduced $\chi^2$ value which is $0.76$, the reconstructed spectrum from the five-year WMAP data 
is $(-)1.1\sigma$ realization. 
In our previous work, the statistical significance of the reconstructed spectrum 
by the cosmic inversion method was estimated to be $(-)1.5\sigma$ in the wave number range from 
$kd\simeq120$ (specifically the band next to the anomalous peak) to $kd\simeq380$. In the same wave number range, 
the statistical significance of the reconstructed spectrum by the maximum likelihood reconstruction method is 
found to be $(-)1.7\sigma$. 
The cumulative deviation from the best-fit power-law spectrum is quite small especially on the scales 
of $kd \gtrsim 130$. 

In our previous analysis, we found a hint of fine structure around $kd \simeq 120$, 
or equivalently the length scale of $\approx 700$Mpc, through the reconstruction by the cosmic inversion method. 
Here, also by the maximum likelihood reconstruction method, 
we find the prominent deviation from the best-fit power-law model on the corresponding scale, 
however, its statistical significance turn out to be a $3.3\sigma$ level. 
Since the statistical distribution of band power is Gaussian, the probability that 
a band power exhibits $3.3\sigma$ or larger excess is $\sim 0.05$\%. 
Indeed, among $50000$ mock samples, we find $31$ realizations giving the same or larger excess 
on the corresponding scale. The probability for realizing a 
primordial spectrum including such excess is about $2$\%, which is simply estimated as the product of 
the one band probability and the number of independent bands. 
This deviation itself is uncommon but not too rare. The reconstructed spectrum is marginally acceptable 
as a realization of a power-law spectrum. 

\begin{figure}
\includegraphics[scale=1.1]{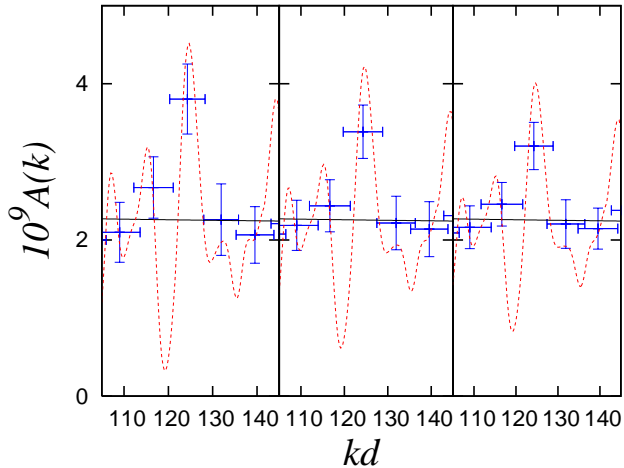}
\caption{\baselineskip 0.35cm
The dependence of reconstruction on the smoothness prior parameter $\epsilon$. 
From left to right, $\epsilon = 3 \times 10^{-4}$Mpc$^{-1}$, 
$4 \times 10^{-4}$Mpc$^{-1}$, and $5 \times 10^{-4}$Mpc$^{-1}$.
}
\label{fig:epsil}
\end{figure}

\begin{figure}
\includegraphics[scale=1.1]{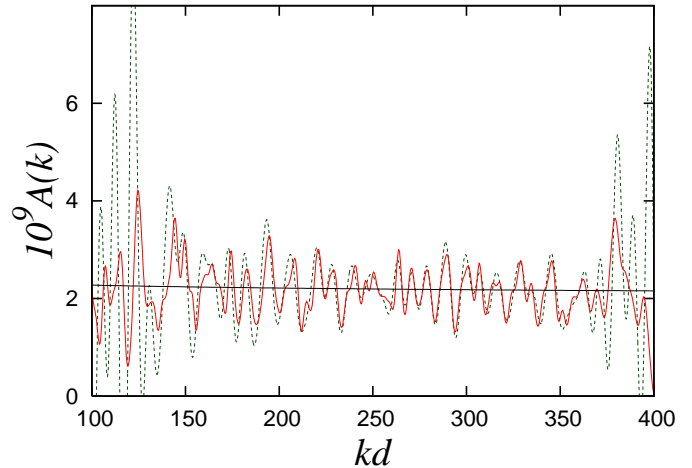}
\caption{\baselineskip 0.35cm
The comparison of the results of two different reconstruction methods. 
The sold curve is the reconstructed primordial spectrum by 
the maximum likelihood reconstruction method, where the prior parameter $\epsilon$ is set to $4 \times 10^{-4}$Mpc$^{-1}$. 
The dotted curve is that by the cosmic inversion method. 
}
\label{fig:compare}
\end{figure}

\section{Conclusion}

We have reconstructed the primordial power spectrum of curvature fluctuation 
by the maximum likelihood reconstruction method, assuming the absence of 
isocurvature modes and the best-fit values of cosmological 
parameters for the power-law $\Lambda$CDM model. 
In a wide range of scales, where the cosmic variance currently dominates over measurement errors ($30 \lesssim kd \lesssim 390$), 
we can reproduce the fine structures 
of modulations off a simple power-law with which we can recover 
the highly oscillatory features observed in $C_\ell$. 

The statistical significance of the oscillatory structure in the
reconstructed power spectrum is difficult to quantify due to the
strong correlation among the neighboring wave number modes. We have therefore 
performed the covariance matrix analysis to calculate the window matrix, 
which diagonalizes the covariance matrix into statistically independent modes. 

In regard to the cumulative deviation ($\chi^2$), 
the reconstructed spectrum is a common realization of a power-law spectrum. 
In particular, on the scales of $kd \gtrsim 130$, the best-fit power-law spectrum for the five-year WMAP data 
fits the reconstructed spectrum slightly excessively well. 
We have focused on the structure on the scale of $kd \simeq 120$, where previously we found the prominent deviation 
from the power-law prediction. 
Although the corresponding band exhibits some deviation from a power-law spectrum, 
we have found that the feature is marginally consistent with a power-law model. 

Finally, we briefly comment on reconstruction from the polarization angular spectrum. 
The WMAP temperature-polarization cross correlation spectrum exhibits 
remarkable deviation from the power-law prediction around $\ell \simeq 120$ \cite{LAMBDA}. 
The enhancement of (negative) correlation observed there seems to agree with 
the anomalous peak found in the reconstructed primordial spectrum in this work, and besides 
they have apparently similar coherent width. 
Although the current polarization data is still suffering large measurement errors, 
the reconstruction from the precise polarization data, which will be available in near future \cite{PLANCK},  
promisingly brings it to a conclusion. 

\acknowledgements
We would like to thank K. Ichiki for discussions. 
We are also grateful to F. Bouchet, D. Spergel for useful comments. 
This work was partially supported by JSPS Grant-in-Aid for Scientific 
Research No. 19340054(JY), JSPS-CNRS Bilateral Joint
Project ``The Early Universe: A Precision Laboratory for High Energy
Physics,'' and JSPS Core-to-Core Program ``International Research
 Network for Dark Energy.'' 

\appendix*
\section{Dependence on the smoothness prior parameter}
To confirm the robustness of our conclusion, it is important to check its 
dependence on the prior parameter $\epsilon$. 
We repeated the statistical analysis described in Sec. \ref{wind} and Sec. \ref{BPA} with different values of $\epsilon$. 
As depicted in Fig. \ref{fig:epsil}, smaller value of $\epsilon$ results in larger amplitude of oscillation 
and then the expected dispersion of the corresponding band power also increases accordingly. 
The error bars evaluated from the statistics of mock samples contains the contribution of 
the systematic effect, which comes from the dependence on the value of $\epsilon$. 
The statistical significance of the anomalous peak concerned in this paper 
lies around $3.3\sigma \sim 3.5\sigma$ from the best-fit power-law spectrum. 
In this regard, our reconstruction scheme is not sensitive to $\epsilon$. 

Since we are motivated by the result of reconstruction by the cosmic inversion method \cite{NY08}, 
it is appropriate to choose the value of $\epsilon$, which gives the similar primordial spectrum 
as that by the cosmic inversion method in the wave number domain where the reconstruction by 
the cosmic inversion method is accurate. We set the value of $\epsilon$ to $4 \times 10^{-4}$Mpc$^{-1}$. 
Figure \ref{fig:compare} shows the reconstructed primordial spectra from the five-year WMAP data by 
the cosmic inversion method and the maximum likelihood reconstruction method. 
The spectrum reconstructed by the maximum likelihood reconstruction method fits the one by the cosmic inversion method 
except on the scales where the cosmic inversion method is inaccurate. 
In this choice of $\epsilon$, we can directly compare the results of the two different reconstruction methods, and 
besides the power of the reconstructed primordial spectrum by the maximum likelihood reconstruction method is 
positive on any scale concerned in this paper.

\end{document}